\documentclass[doublecol]{epl2}
\usepackage{amsmath,amssymb} 
\usepackage{graphicx,color,bm}
\usepackage{cite}
\usepackage{subfigure}

\DeclareMathOperator{\tr}{tr}
\DeclareMathOperator{\rank}{rank}
\DeclareMathOperator{\sign}{sgn}
\DeclareMathOperator{\diag}{diag}

\providecommand{\eff}[0]{e}
\providecommand{\mean}[1]{\langle #1 \rangle}
\providecommand{\bra}[1]{\langle #1 |}
\providecommand{\iprod}[2]{\langle #1 |#2 \rangle}
\providecommand{\ket}[1]{| #1 \rangle}
\providecommand{\dotop}[0]{{\bm\cdot\,}}
\providecommand{\VBS}[0]{\text{VBS}}
\providecommand{\MPS}[0]{\text{MPS}}
\providecommand{\KSZ}[0]{\text{KSZ}}

\providecommand{\fabs}[1]{\left| #1 \right|}

\def\vec#1{\text{\bfseries#1}}
\def\mat#1{\text{\sffamily\bfseries#1}}

\title{Entanglement spectra of the $q$-deformed Affleck-Kennedy-Lieb-Tasaki model and matrix product states}
\shorttitle{Entanglement spectra of the $q$-deformed AKLT model and MPS} 

\author{R. A. Santos\inst{1} \and F. N. C. Paraan\inst{1} \and V. E. Korepin\inst{1} \and A. Kl\"umper\inst{2}}
\shortauthor{R. A. Santos \etal}

\institute{                    
  \inst{1} C. N. Yang Institute for Theoretical Physics, State University of New York at Stony Brook - New York 11794-3840, USA\\
  \inst{2} Fachbereich C Physik, Bergische Universit\"at Wuppertal - 42097 Wuppertal, Germany
}

\abstract{
We exactly calculate the reduced density matrix of matrix product states (MPS). Our compact result enables one to perform analytic studies of entanglement in MPS. In particular, we consider the MPS ground states of two anisotropic spin chains. One is a $q$-deformed Affleck-Kennedy-Lieb-Tasaki (AKLT) model and the other is a general spin-1 quantum antiferromagnet with nearest-neighbor interactions. Our analysis shows how anisotropy affects entanglement on different continuous parameter manifolds. We also construct an effective boundary spin model that describes a block of spins in the ground state of the $q$-deformed AKLT Hamiltonian. The temperature of this effective model is given in terms of the deformation parameter $q$.
}

\begin{document}

\maketitle

\section{Introduction}
We study entanglement in the ground state of spin models. Specifically, we consider spin-1 chains with nearest-neighbor interactions. We begin by analyzing the reduced density matrix $\rho_\ell$ of a block of sequential spins within the chain. The reduced density matrix is fundamental in the study of entanglement. It is used to define the R\'enyi and von Neumann entanglement entropies
\begin{align}
	S_\text{R}(\alpha) &\equiv \frac{\log\tr\rho_\ell^\alpha}{1-\alpha},\\
	S_\text{vN} &\equiv -\tr(\rho_\ell\log\rho_\ell) = \lim_{\alpha\to 1} S_\text{R}(\alpha).
\end{align}	
These quantities are frequently used measures of entanglement \cite{bennett1996a,bennett1996c,vedral1997,amico2008}. Also, the eigenvalues of $\rho_\ell$ are used to construct the entanglement spectrum of the block \cite{li2008,calabrese2008,lou2011,cirac2011}. This spectrum describes the mixed state of the block as if it were governed by an effective Hamiltonian at some temperature.

In this paper, the reduced density matrix of translationally invariant matrix product states (MPS) \cite{fannes1989,verstraete2006,verstraete2008,aklt1987,aklt1988,kirillov1989,totsuka1995,karimipour2008,klumper1991,klumper1993} is calculated exactly. This class of MPS includes one-dimensional valence-bond-solid (VBS) states \cite{aklt1987,aklt1988,kirillov1989,totsuka1995,karimipour2008} and ground states of anisotropic spin chains \cite{klumper1991,klumper1993}.
In particular, we consider the MPS ground states $\ket{\MPS}$ of two nonequivalent anisotropic generalizations of the Affleck-Kennedy-Lieb-Tasaki (AKLT) model \cite{aklt1987,aklt1988}. Both are spin-1 chains described by Hamiltonians of the form $\mathcal{H} = \sum_j h_{j,j+1}$. The local Hamiltonian $h_{j,j+1}$ acts on the Hilbert space of neighboring spins $j$ and $j+1$. Additionally,
it annihilates the frustration-free ground states of these models, $h_{j,j+1}\ket{\MPS} = 0$. 

The first generalization we consider is a $q$-deformed AKLT (AKLT$_q$) model that is $\text{SU}_q$(2) invariant \cite{batchelor1990,klumper1991}. The second model is a general Hamiltonian with U(1) symmetry introduced by Kl\"umper, Schadschneider, and Zittartz (KSZ model) \cite{klumper1993}. The KSZ Hamiltonian is invariant under (i) lattice translations and reflections, (ii) spin rotations about the longitudinal $z$-axis, and (iii) spin reflections about the transverse plane $S_i^z \to -S_i^z$. We discuss each in more detail in the following subsections.

Entanglement in the VBS ground state of isotropic AKLT models has been studied in the literature \cite{fan2004,katsura2007,xu2008,katsura2008,geraedts2010,orus2011}. However, the question of how anisotropy modifies entanglement in these states has received less attention \cite{verstraete2004a,jin2004,chen2006}. 
We address this matter by calculating the entanglement spectra and entropies in AKLT$_q$ and KSZ ground states as functions of anisotropy parameters. 

\section{Reduced density matrix of MPS}
We start with a pure matrix product state $\ket{\MPS}$. The density matrix of the whole state is $\rho = \ket{\MPS}\bra{\MPS}/\iprod{\MPS}{\MPS}$. We then partition the system into a block of $\ell$ sequential spins and its environment $E$. The reduced density matrix $\rho_\ell$ is the partial trace of $\rho$ over the environment, $\rho_\ell = \tr_E \rho$.

The MPS representation of a periodic chain of $L$ identical spins is
\begin{equation}\label{mps}
	\ket{\MPS} = \tr[\mat{g}_1\dotop \mat{g}_2 \dotop \negthinspace\dotsc \dotop \mat{g}_L].
\end{equation}
The $\mat{g}_j$ are $D\times D$ matrices ($D=2$ for the examples considered below). The trace here is taken over the auxiliary matrix space (not $E$). The elements of $\mat{g}_j$ are
\begin{equation}
	(\mat{g}_j)_{\alpha\beta} = \sum_{m} A_{\alpha\beta}{(m)}\ket{m}_j.
\end{equation}
The set $\{\ket{m}_j\}$ is a complete orthonormal basis for the Hilbert space of the spin at site $j$ and the coefficients $A_{\alpha\beta}{(m)}$ are independent of the site index. Due to translational invariance, we drop the site label $j$ whenever possible. We denote the matrix dual to $\mat{g}$ as $\bar{\mat{g}}$ with elements $(\bar{\mat{g}})_{\alpha\beta} = \sum_{m} A_{\alpha\beta}^*{(m)}\bra{m}$. Here, the coefficients are replaced by their complex conjugates and the kets are replaced by the corresponding bras. In Eq.~\eqref{mps} the matrix multiplication $({\bm\cdot})$ involves tensor products of vector matrix elements, i.e.,
\begin{align}
	(\mat{g}_j\dotop\mat{g}_{j+1})_{\alpha\gamma} = \sum_{\beta mn}A_{\alpha\beta}{(m)}A_{\beta\gamma}{(n)}\ \ket{m}_{j}\otimes \ket{n}_{j+1}.
\end{align}
The dual $(\bar{\mat{g}}_j\dotop\bar{\mat{g}}_{j+1})_{\alpha\gamma}$ is defined analogously. For products of $\mat{g}$ matrices denoting a block of sequential spins we introduce an abbreviation
\begin{equation}\label{blockstate}
	(\mat{g}_j\dotop \mat{g}_{j+1} \dotop \negthinspace\dotsc \dotop \mat{g}_{j'})_{\alpha\alpha'} = \ket{\alpha\alpha';j,j'}.
\end{equation}
Thus, it is sufficient to identify the states of boundary spins to specify the state of a block. 

Let us construct a transfer matrix $\mat{G}\equiv\bar{\mat{g}}\otimes\mat{g}$ that is useful for calculating state overlaps (scalar products) and correlation functions:
\begin{equation}\label{transfer}
	(\mat{G})_{\alpha\gamma,\beta\delta} = (\bar{\mat{g}})_{\alpha\beta}(\mat{g})_{\gamma\delta} = \sum_{m} A_{\alpha\beta}^*(m)A_{\gamma\delta}(m).
\end{equation}
For example, the square of the norm of  $\ket{\MPS}$ is $\iprod{\MPS}{\MPS} = \sum_{\alpha\alpha'}\iprod{\alpha\alpha';1,L}{\alpha\alpha';1,L} = \tr \mat{G}^L$.

The density matrix is therefore
\begin{equation}
	\rho =  \frac{\tr[\mat{g}_1\dotop\negthinspace\dotsc\dotop \mat{g}_L] \tr[\bar{\mat{g}}_1\dotop\negthinspace\dotsc\dotop \bar{\mat{g}}_L]}{\tr \mat{G}^L},
\end{equation}
while the reduced density matrix is
\begin{align}\label{rho1}
	\rho_\ell = \sum_{\alpha\alpha'\beta\beta'}\frac{\ket{\alpha\alpha';1,\ell}	\bigl(\mat{G}^{L-\ell}\bigr)_{\alpha'\beta',\alpha\beta} \bra{\beta\beta';1,\ell}}{\tr\mat{G}^L}.
\end{align}
We chose the block to extend from site 1 to site $\ell$ without loss of generality because of translational invariance. 
In this form, $\rho_\ell$ clearly acts nontrivially only in the subspace spanned by the block state vectors $\{\ket{\alpha\alpha';1,\ell}\}$. The dimension of this subspace is at most $D^2$.

Let us consider transfer matrices with the symmetries $(\mat{G})_{\alpha\beta,\alpha'\beta'} =(\mat{G})_{\alpha'\beta',\alpha\beta}  = (\mat{G})_{\beta\alpha,\beta'\alpha'}$. This is a weak requirement because it means that the scalar product $\iprod{\alpha\alpha';1,n}{\beta\beta';1,n}$ is invariant under lattice reflections (e.g., `flipping' the ring over). We construct a symmetric overlap matrix $\mat{K}(n)$ that is related to the $n\textsuperscript{th}$ power of $\mat{G}$ by ${\bm(}\mat{K}(n){\bm)}_{\alpha\alpha',\beta\beta'} \equiv (\mat{G}^n)_{\alpha\beta,\alpha'\beta'}$.
With this definition we write $\rho_\ell$ as
\begin{align}\label{rhoprime}
	\rho_\ell = \sum_{\alpha\alpha'\beta\beta'} \frac{\ket{\alpha\alpha';1,\ell}	\bigl(\mat{K}(L-\ell)\bigr)_{\alpha\alpha',\beta\beta'} \bra{\beta\beta';1,\ell}}{\tr\mat{G}^L}.
\end{align}
The indices are now matched so that we can express $\rho_\ell$ as a product of matrices. Suitable similarity transformations within the space spanned by $\{\ket{\alpha\alpha';1,\ell}\}$ finally gives
\begin{equation}\label{rhox}
	\rho_\ell = \frac{\mat{K}(L-\ell){\mat{K}}(\ell)}{\tr[\mat{K}(L-\ell){\mat{K}}(\ell)]}. 
\end{equation}

This formula has some notable features. First, it is a general expression that is valid for a large class of MPS.
Also, we find that $\rho_\ell$ has a small number of nonzero eigenvalues, $\rank\rho_\ell \le D^2$ \cite{wolf2006,michalakis2006}.
Furthermore, it is evident that $\rho_\ell$ and $\rho_{L-\ell}$ are isospectral. Thus, the entanglement entropies of the block and environment are the same. 

\section{AKLT$_q$ model}
The AKLT$_q$ Hamiltonian is given in the Appendix. It is hermitian for real $q$ and has the symmetry of the SU$_q$(2) quantum group \cite{drinfeld1985,jimbo1985}. This deformation of SU(2) symmetry appears naturally in the anisotropic spin-1/2 XXZ Heisenberg model \cite{reshetikhin1989}.

The generators of the SU$_q$(2) quantum algebra are $\mathbb{S}^{z,\pm}\equiv\sum_j\mathbb{S}_j^{z,\pm}$. They satisfy the commutation relations
\begin{equation}
[\mathbb{S}^+,\mathbb{S}^-] = \frac{q^{2\mathbb{S}^z} -q^{-2\mathbb{S}^z}}{q-q^{-1}},\quad [\mathbb{S}^z,\mathbb{S}^\pm] = \pm \mathbb{S}^{\pm}.
\end{equation}
The $q$-deformed spin-1 operators at site $j$ are $\mathbb{S}_j^z = S_j^z$ and
\begin{equation}
	\mathbb{S}_j^{\pm} =\sqrt{\frac{q+q^{-1}}{2}}\bigl(q^{-S_1^z}\otimes \dotsb \otimes q^{-S_{j-1}^z} \otimes S_j^{\pm} \otimes q^{S_{j+1}^z} \otimes \dotsb\bigr). \label{deformedspin}
\end{equation}
Here $S_j^{z,\pm}$ are undeformed spin-1 operators at site $j$. Eq.~\eqref{deformedspin} has similarities with Jordan-Wigner transformations.

The entanglement spectrum does not depend on the sign of $q$ and is invariant under the transformation $q\to q^{-1}$. Thus, we will only consider $q \in (0,1]$. The isotropic AKLT model is recovered at $q=1$.

The unique ground state of the periodic AKLT$_q$ model is a $\VBS_q$ state. Its MPS representation was constructed in Ref.~\cite{klumper1991}. The properties of this state and its spin-spin correlation functions have been studied \cite{klumper1991,klumper1992}. This approach was later extended to obtain the spin-spin correlators of higher integer spin-$S$ AKLT$_q$ chains \cite{motegi2010,arita2011}. 

The local Hamiltonian $h_{j,j+1}$ is a {\bf projector} onto the subspace of the $q$-deformed spin-2 quintuplet formed by adjacent spins \cite{klumper1991}. The MPS form of the $\VBS_q$ ground state is obtained by requiring $h_{j,j+1}$ to annihilate the elements of the matrix $\mat{g}_j\dotop\mat{g}_{j+1}$. This condition leads to 
\begin{equation}
	\mat{g} = \begin{pmatrix} q^{-1} \ket{0} & -\sqrt{q+q^{-1}}\ket{+} \\ \sqrt{q+q^{-1}}\ket{-} & -q\ket{0} \end{pmatrix},
\end{equation}
where the vector elements are eigenstates of $S^z$. This yields the MPS representation
\begin{equation}
	\ket{\text{VBS}_q} = \tr[\mat{g}_1\dotop\mat{g}_2\dotop\negthinspace \dotsc\dotop\mat{g}_L].
\end{equation}
Since $\mat{g}$ is a $2\times 2$ matrix, the reduced density matrix has at most four nonzero eigenvalues.

In the limit $q\to 0$, the AKLT$_q$ Hamiltonian is dominated by (classical) Ising-type interactions.  In this case the $\mat{g}$ matrix has only one diagonal element $\ket{0}$. The resulting ground state is a product state $\bigotimes_{j}\ket{0}_j$ describing a magnet polarized in the transverse direction. In this classical limit all spins are in the $S_j^z = 0$ state. Hence, any block in the chain has zero entropy.

\begin{figure}[tb]
\begin{center}
\subfigure[]{
\includegraphics[width=0.48\linewidth]{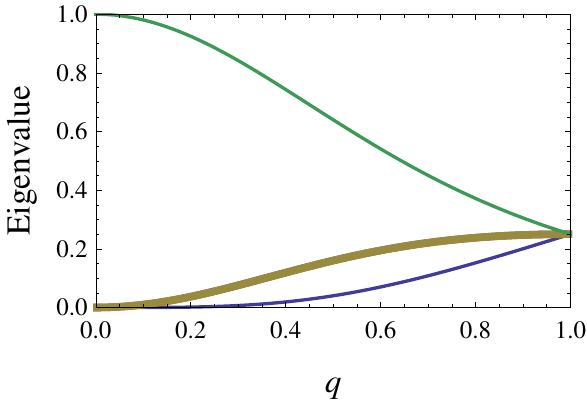}}
\subfigure[]{
\includegraphics[width=0.48\linewidth]{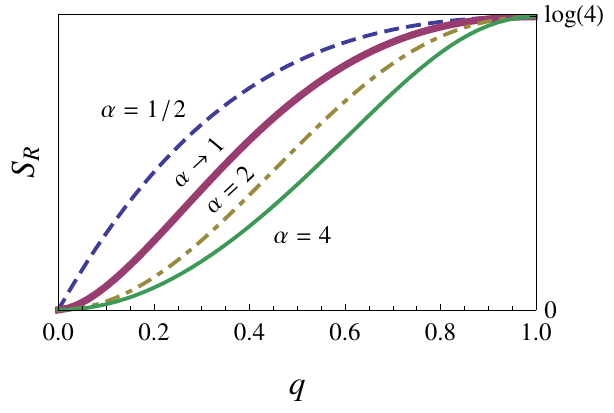}}
\caption{(Color online) (a) In the double scaling limit the reduced density matrix of $\ket{\VBS_q}$ has four degenerate eigenvalues at the isotropic $q=1$ point. The two middle eigenvalues are always degenerate (bold line). In the classical limit $q\to 0$ the ground state is a product state and the only nonzero eigenvalue is unity.  (b) The R\'enyi entropy $S_\text{R}$ vanishes as $q\to 0$. It saturates to the maximally entangled value $S_\text{R} = \log 4$ at the AKLT point $q=1$. The von Neumann entropy (bold line) is obtained at $\lim_{\alpha\to1} S_\text{R}$.\label{qlargeblock}}
\end{center}
\end{figure}

The transfer matrix \eqref{transfer} for the $\ket{\VBS_q}$ state is 
\begin{equation}
	\mat{G}\equiv \bar{g} \otimes g =  \begin{pmatrix} q^{-2}& 0& 0& q+q^{-1}\\ 
								0& -1& 0& 0\\
								 0& 0& -1& 0\\
							 q+q^{-1}& 0& 0& q^2\end{pmatrix}.
\end{equation}
The eigenvalues of this matrix are $\{\Lambda,-1,-1,-1\}$ and the dominant eigenvalue is $\Lambda = 1+q^2+q^{-2} \ge 3$. The overlap matrix $\mat{K}(\ell)$ is therefore
\begin{multline}
	{\mat{K}(\ell)} = \frac{\Lambda^\ell}{q+q^{-1}}\,\diag{\{q^{-1},1,1,q\}} \\ + \frac{(-1)^\ell}{q+q^{-1}}\begin{pmatrix} q & 0 &0 & q+q^{-1}\\
	0 & -1 &0 & 0\\
	0 & 0 &-1 & 0\\
	q+q^{-1} & 0 &0 & q^{-1} \end{pmatrix}.
\end{multline}
Eq.~\eqref{rhox} gives the reduced density matrix
\begin{equation}
	\rho_\ell = \frac{\mat{K}(L-\ell)\mat{K}(\ell)}{\Lambda^L+3(-1)^L}.
\end{equation}

In the double scaling limit of an infinite block $\ell\to\infty$ in an infinite chain $(L-\ell)\to\infty$, the reduced density matrix becomes diagonal. Thus, the block states $\ket{\alpha\alpha';1,\ell}$ are orthogonal to each other. The eigenvalues $\{p_i\}$ of the reduced density matrix $\rho_{\ell\to\infty}$ are
\begin{equation}
  p_{1,4} = \frac{q^{\mp 2}}{(q+q^{-1})^2}, \ \ p_{2}= p_{3}= \frac{1}{(q+q^{-1})^2}, \ \ell\to\infty.
\end{equation}
We discover an important consequence of $q$-deformation: The degeneracy of the entanglement spectrum changes between the classical and isotropic AKLT points. This result is depicted in Fig.~\ref{qlargeblock}(a).

To obtain an intuitive picture for the mixed state of the block we write $\rho_{\ell\to\infty} = \text{e}^{-\beta H_\eff}/\tr \text{e}^{-\beta H_\eff}$. $H_\eff$ is an effective Hamiltonian and $1/\beta$ an effective temperature. The eigenvalues of $H_\eff$ form the entanglement spectrum \cite{li2008}. Doing so gives the effective temperature $1/\beta = 1/\fabs{\ln q}$ and effective Hamiltonian
\begin{equation}
	H_\eff = \sigma_1^x\sigma_\ell^x + \sigma_1^y\sigma_\ell^y,
\end{equation}
where $\sigma_j^i$ are Pauli operators at site $j$. Thus, $\rho_{\ell\to\infty}$ describes a thermal ensemble of two spin-1/2's at the block boundaries with Heisenberg (XX) interaction. The anisotropy parameter $q$ determines the effective boundary temperature $T_\eff = 1/\fabs{\ln q}$. For the original AKLT model ($q=1$) the effective boundary spins are in a maximally mixed state ($T_\eff\to\infty$), while in the classical limit $q\to0$ they are in a pure state ($T_\eff =0$).

This result for the effective Hamiltonian is consistent with the area law for gapped models \cite{hastings2007,eisert2010}. It is similar to the effective boundary spin chain proposed for 2D AKLT models \cite{katsura2010,lou2011}. However, in the AKLT$_q$ chain the effective boundary spin interaction is long-ranged and exists for arbitrarily long blocks. The long range of this interaction follows from the non-local nature of the SU$_q$(2) symmetry \eqref{deformedspin} of the model.

The R\'enyi and von Neumann entropies in the double scaling limit are
\begin{align}
	S_\text{R}^{\ell\to\infty} &= \frac{2}{1-\alpha}\log\frac{q^\alpha+q^{-\alpha}}{(q+q^{-1})^\alpha}, \\
	S_\text{vN}^{\ell\to\infty} &= \log(q+q^{-1})^2-\frac{q-q^{-1}}{q+q^{-1}}\log q^2.
\end{align}
We plot these entropies in Fig.~\ref{qlargeblock}(b) and observe the effects of $q$-deformation even for infinitely long blocks. Anisotropy reduces entanglement entropy from its maximum value $S_\text{R} = \log 4$. This maximum is reached at the AKLT point \cite{fan2004}. 

\begin{figure}[tb]
\begin{center}
\subfigure[]{
\includegraphics[width=0.48\linewidth]{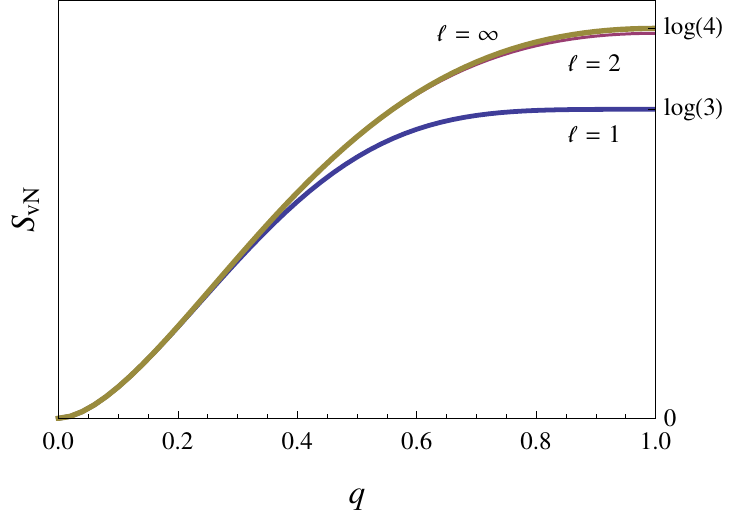}}
\subfigure[]{
\includegraphics[width=0.48\linewidth]{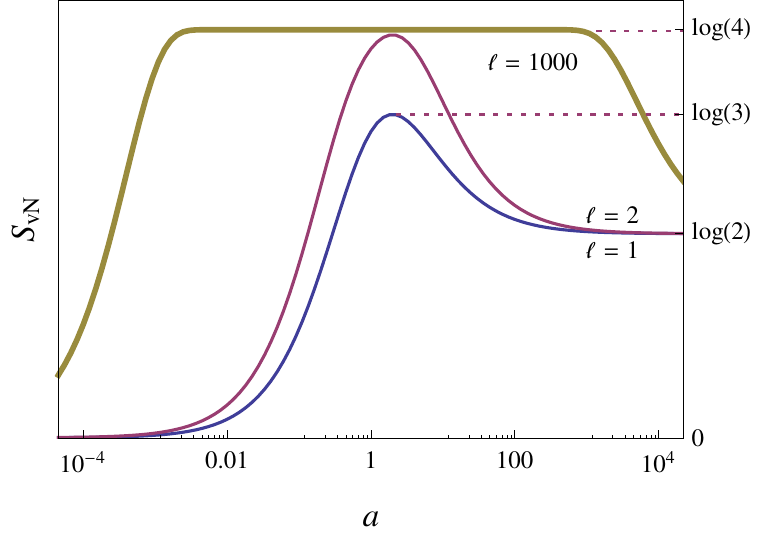}}
\caption{(Color online) The von Neumann entropy decreases away from the isotropic AKLT points $q=1$ for $\ket{\VBS_q}$ (a) and $a=2$ for $\ket{\KSZ_a}$ (b). At the isotropic point finite-size corrections are largest for $\ket{\VBS_q}$ and smallest for $\ket{\KSZ_a}$. \label{qblock}}
\end{center}
\end{figure}

For blocks of finite length $\ell < \infty$ in an infinite chain $L\to\infty$, the overlap matrix $\mat{K}(\ell)$ has off-diagonal terms. In this case, the eigenvalues of $\rho_\ell$ acquire finite-size corrections:
\begin{align}
p_{1,4}&=\frac{q^2+q^{-2}+2(-\Lambda)^{-\ell}}{2(q+q^{-1})^2}\pm\sqrt{\frac{1}{4}-\frac{1-(-\Lambda)^{-2\ell}}{(q+q^{-1})^2}}, \nonumber \\
p_{2}&=p_{3}=\frac{1-(-\Lambda)^{-\ell}}{(q+q^{-1})^2}, \quad \Lambda = 1+q^2+q^{-2}. \label{qvbseigen} 
\end{align}
These eigenvalues are exact. The corrections decay exponentially as expected for a gapped system. At $q = 1$ we recover the eigenvalues 
\begin{equation}
	p_1 = \frac{1+3(-3)^{-\ell}}{4}, \quad p_2 = p_3=p_4 = \frac{1-(-3)^{-\ell}}{4},
\end{equation}
for the isotropic AKLT chain \cite{fan2004}. Figure \ref{qblock}(a) shows how $q$-deformation reduces the von Neumann entropy. For a block consisting of one spin ($\ell=1$) one eigenvalue is identically zero. Thus, the maximum single-site von Neumann entropy is $\log 3$. This value is reached at the isotropic point $q=1$ where there is a uniform mixture of three spin-1 states.

For large but finite blocks $1\ll \ell<\infty$ and $q\ne1$, an expansion of the eigenvalues \eqref{qvbseigen} gives the leading order corrections $\pm (-1)^{\ell}(q+q^{-1})^{-2}\text{e}^{-\ell/\xi}$. The characteristic length of these corrections is $\xi =1/\ln(1+q^2+q^{-2})$. This quantity is equal to the correlation length of the spin-spin correlators of the VBS$_q$ state \cite{klumper1991,klumper1992,arita2011}.

\section{KSZ model}
The local KSZ Hamiltonian is given by
\begin{multline}
	h_{j,j+1} = \alpha_0A_j^2 +\alpha_1(A_jB_j+B_jA_j) + \alpha_2B_j^2+\alpha_3A_j \\ + \alpha_4B_j(1+B_j) + \alpha_5\bigl[(S_j^z)^2+(S_{j+1}^z)^2\bigr] + \alpha_6,
\end{multline}
with a transverse interaction term $A_j = S_j^xS_{j+1}^x +S_j^yS_{j+1}^y$, longitudinal interaction term $B_j = S_j^zS_{j+1}^z$, and constants $\alpha_i$. Requiring $h_{j,j+1}$ to have nonnegative eigenvalues and annihilate an MPS ground state $\ket{\KSZ_a}$ leads to a submanifold of Hamiltonians with restrictions on the constants $\alpha_i$ \cite{klumper1993}. The correlation functions and low-lying excitations of this model have been studied \cite{klumper1993,bartel2003}, but its entanglement spectrum has not yet been investigated.

We obtain the MPS form of $\ket{\KSZ_a}$ from the $\mat{g}$ matrix
\begin{equation}
	\mat{g} = \begin{pmatrix} \ket{0}&-\sqrt{a}\ket{+} \\ \sqrt{a}\ket{-}& -\sigma\ket{0}\end{pmatrix},
\end{equation}
where $a>0$ is an anisotropy parameter and $\sigma = \sign\alpha_3$.  The corresponding transfer matrix is
\begin{equation}
	\mat{G} = \begin{pmatrix} 1& 0& 0& a\\
																															0& -\sigma & 0& 0\\ 
																															0& 0& -\sigma & 0\\
																															a& 0& 0& 1\end{pmatrix}.
\end{equation}
The unique ground state is $\ket{\KSZ_a} = \tr[\mat{g}_1\dotop\mat{g}_2\dotop\negthinspace \dotsc\dotop\mat{g}_L]$ (periodic boundary conditions). It reduces to the isotropic VBS state at $a=2$ and $\sigma=1$. 

The eigenvalues of $\rho_\ell$ for an infinite chain $L\to \infty$ are
\begin{align}
	p_{1,4} &= \frac{1}{4}\biggl[1+ \biggl(\frac{1-a}{1+a}\biggr)^\ell \pm 2\biggl(\frac{-\sigma}{1+a}\biggr)^\ell \biggr], \nonumber\\
	p_2 &= p_3 = \frac{1}{4}\biggl[1- \biggl(\frac{1-a}{1+a}\biggr)^\ell \biggr]. \label{kszeigen}
\end{align}
We observe that the entanglement spectrum is the same for $\sigma = \pm 1$. In the double scaling limit the eigenvalues of $\rho_{\ell\to\infty}$ are four-fold degenerate $p_i^{\ell\to\infty} = \tfrac{1}{4}$. The block is maximally entangled with $S_\text{R}=\log4$. The entanglement spectrum therefore corresponds to a four-level system at infinite temperature.

Let us now consider blocks of finite length. The von Neumann entropy is a maximum at the isotropic point $a=2$. This property is depicted in Fig.~\ref{qblock}(b). For a block of one spin ($\ell =1$) one eigenvalue of $\rho_\ell$ vanishes and the maximum entanglement entropy is $\log 3$. 
In the limit $a\to 0$, the $\ket{\KSZ_a}$ ground state approaches the transverse ferromagnet $\bigotimes_{j}\ket{0}_j$. This is a (classical) product state with no entanglement. In the opposite limit $a\to\infty$ the reduced density matrix represents a uniform mixture of two degenerate N\'eel ordered states. In this limit the von Neumann entropy approaches $\log 2$. 

Finite-size corrections to the eigenvalues \eqref{kszeigen} are exponential in $\ell$. The characteristic lengths of these corrections are $\xi_\| = 1/\ln\fabs{(1+a)/(1-a)}$ and $\xi_\bot = 1/\ln(1+a)$. These quantities are equal to the longitudinal ($\xi_\|$) and transverse ($\xi_\bot$) correlation lengths of the spin-spin correlation functions \cite{klumper1993}:
\begin{align}
	{\mean{S_1^zS_\ell^z}} &= -\frac{a^2}{(1-a)^2}[\sign(1-a)]^{\ell}\times \text{e}^{-\ell/\xi_\|},\\
	{\mean{S_1^xS_\ell^x}} &= -a (\sigma + 1)[\sign(-\sigma)]^\ell\times \text{e}^{-\ell/\xi_\bot}, \quad \ell \ge 2.
\end{align}


\section{Conclusions}
In this paper we studied entanglement in the MPS ground states of the anisotropic AKLT$_q$ and KSZ models. We derived a compact formula for the reduced density matrix in terms of the MPS transfer matrix. With this result we calculated entanglement entropies on continuous parameter spaces connecting the isotropic VBS state with classical states. 

In the AKLT$_q$ model we found that $q$-deformation reduces entanglement in the $\ket{\VBS_q}$ state even in the double scaling limit. Our analysis enabled us to construct an effective Heisenberg model for the boundary spins of infinitely long blocks. This important result identifies the deformation parameter $\left|\ln q\right|^{-1}$ as an effective temperature for these boundary spins.

Anisotropy in the KSZ model also reduces entanglement in the $\ket{\KSZ_a}$ state for blocks of finite length. But unlike the $\ket{\VBS_q}$ state, the boundary spins of infinitely long blocks are maximally entangled at fixed anisotropy $0<a<\infty$. Thus, the effective boundary spins in the double scaling limit are at infinite temperature.

\acknowledgments
The authors are grateful for the hospitality of the Simons Center for Geometry and Physics (Stony Brook, New York). R.S., F.P., and V.K. acknowledge support by the National Science Foundation through Grant No. DMS-0905744. R.S. is supported by a Fulbright-CONICYT grant. A.K. thanks the C.~N.~Yang Institute for Theoretical Physics for its hospitality. 

\section{Appendix}\label{appendix}
The AKLT$_q$ Hamiltonian is
\begin{align}\nonumber 
	\mathcal{H} &= b\,{\sum_{j}}\, \bigl\{  c\,\vec{S}_j\cdot\vec{S}_{j+1} + \bigl[\vec{S}_j\cdot\vec{S}_{j+1} \nonumber \\ 
	&\qquad\quad + \tfrac{1}{2}(1-c)(q+q^{-1}-2)S_j^zS_{j+1}^z \nonumber \\ 
	&\qquad\quad + \tfrac{1}{4}(1+c)(q-q^{-1})(S_{j+1}^z-S_{j}^z)\bigr]^2 \nonumber \\ 
	&\quad + \tfrac{1}{4}\,c\,(1-c)(q+q^{-1}-2)^2(S_j^zS_{j+1}^z)^2 \nonumber \\ 
	&\quad + \tfrac{1}{4}\,c\,(1+c)(q-q^{-1})(q+q^{-1}-2)S_j^zS_{j+1}^z \nonumber \\
	&\qquad\quad \times  (S_{j+1}^z-S_j^z) \nonumber \\
	&\quad +  \tfrac{1}{4}(c-3)\bigl[\bigl(c-1+\tfrac{1}{2}(1+c)^2\bigr)S_j^zS_{j+1}^z  \nonumber \\ 
	&\qquad\quad + 2 \bigl( c- \tfrac{1}{8}(1+c)^2\bigr)\bigl((S_{j+1}^z)^2+(S_j^z)^2\bigr)\bigr] \nonumber \\
	&\quad + (c-1) + \tfrac{1}{2}\,c\,(q^{2}-q^{-2})(S_{j+1}^z-S_j^z)\bigr\},
\end{align}
with $c= 1 + q^{2} +q^{-2}$ and $b=[c\,(c-1)]^{-1}$ \cite{batchelor1990,klumper1992}.


\end{document}